\documentclass[twocolumn,amsmath,amssymb,prl,10pt,nofootinbib,superscriptaddress]{revtex4}
\usepackage{ graphicx, float,amsmath, amsmath, amssymb}
\usepackage[export]{adjustbox}

\def\be{\begin{equation}}
\def\ee{\end{equation}}
\def\bea{\begin{eqnarray}}
\def\eea{\end{eqnarray}}
\def\bse{\begin{subequations}}
\def\ese{\end{subequations}}

\usepackage[breaklinks, colorlinks, citecolor=blue]{hyperref}
\usepackage[labelsep=period]{caption}
\usepackage[normalem]{ulem}
\usepackage{soul}
\usepackage{minitoc}

\usepackage{bm}

\begin{document}
\title{Editorial to the special issue ``probing new physics with black holes"}


\author{Aur\'elien Barrau}%
\affiliation{%
Laboratoire de Physique Subatomique et de Cosmologie, Universit\'e Grenoble-Alpes, CNRS/IN2P3\\
53, avenue des Martyrs, 38026 Grenoble cedex, France
}

\date{\today}
\begin{abstract} 
Black holes are fantastic laboratories to probe new physics. Both theoretically and experimentally, many new ideas are emerging to use them as tools for understanding better quantum gravity or classical gravity beyond general relativity. I briefly review some new results.
 \end{abstract}
\maketitle

Black holes (BHs) are now observed astronomical objects. In a few years only, they have been investigated both thanks to gravitational waves \cite{LIGOScientific:2018mvr,Abbott:2016blz} and with interferometric imaging \cite{Akiyama:2019cqa,Akiyama:2019eap}. Although the idea goes back to the eighteenth century with John Michell and Pierre-Simon de Laplace, we have recently experienced a huge acceleration in black hole physics. Beyond the historical evidences -- accretion discs, jets, orbits of stars \cite{Gillessen:2008qv}, etc. -- we now have a kind of a direct access\footnote{Actually, ``direct" is a misleading concept: strictly speaking, no observation is direct.} to the event horizon.\\

As argued in \cite{Giddings:2019jwy}, inconsistencies between the evolution of BHs and the principles of quantum mechanics, taking into account the Hawking effect, strongly motivate the modification of the classical description of a BH at horizon scales. This means that interesting non-trivial effects could be observed in the images produced by the Event Horizon Telescope (EHT). Combining reasonable hypotheses to remain maximally consistent with known physics, this leads to interactions described by an effective Hamiltonian
\begin{equation}
H_I = \int \sqrt{-g_{tt}}\,dV H^{\mu\nu}(x) T_{\mu\nu}(x)\ .
\end{equation}
Here $dV$ is the volume element on the time slices, $g_{tt}$ is the metric in slice-time $t$, $T_{\mu\nu}(x)$ is the stress-energy tensor for field excitations, and $H^{\mu\nu}(x)$ parameterizes the interactions with the BH. The size of  $H^{\mu\nu}$ is constrained by unitarity. Running this model under a few natural assumptions leads to the prediction of a temporal variability of images. For the M87 BH, the calculated typical time-scale of the variability could potentially be experimentally probed. The period, in days, is given by 
\begin{equation}
P\simeq  59 \left(\frac{M}{6.5\times 10^9 M_\odot}\right)\left({1\over 2} + {1\over 2\sqrt{1-a^2}}\right),
\end{equation}
for an angular momentum $L=aGM$.
\\

The other main experimental axis is the one based on gravitational waves, as now routinely measured by LIGO/Virgo. A promising way to detect subtle effects beyond general relativity relies in the investigation of quasinomal modes (QNMs). They typically correspond to the ringdown phase of a BH merger and are now observed beyond the fundamental mode. As explained in \cite{Moulin:2019ekf}, when considering a  general metric of the form 
\begin{equation}
ds^2=A(r)dt^2-B(r)^{-1}dr^2-H(r)d \theta ^2 - H(r) \sin ^2 \theta
d \phi ^2,
\label{metric}
\end{equation}
the radial equation for the perturbation $R(r)$ reads 
\begin{equation}
H^2 \sqrt{\frac{  B}{A}} \frac{\partial }{ \partial r} \bigg( \frac{\sqrt{AB}}{H } \frac{\partial R(r)}{\partial r} \bigg) + \bigg( \frac{  H}{A} \omega ^2 - \mu ^2 \bigg) R(r)= 0,
\end{equation}
where $w$ is the pulsation and $\mu ^2= (l-1)(l+2)$ for a multipolar momentum $l$. The QNMs have been investigated in massive gravity, STV gravity, Ho\v{r}ava-Lifshitz gravity, quantum corrected gravity, and loop quantum gravity in the WKB approximation. The effects of modifying the gravitational theory were shown to be more important for the real part (frequency) than for the imaginary part (amplitude) of the complex frequency. The sign of the frequency shift, and the way it depends upon the overtone and multipole numbers are characteristic of a given modification of GR. This establishes a new way to experimentally investigate modified gravity.\\

On the theoretical side, an outstanding question of black hole physics lies in the derivation of a full quantum gravity description. A consensual such theory is still missing. However, generic results for polymeric black holes are derived in \cite{Aruga:2019dwq}. The idea is to rely on effective deformations of the phase space of spherically symmetric GR. One works with a deformed Hamiltonian constraint, keeping the usual vectorial constraint, which is expected to lead to a notion of generalized covariance. But, using a Lagrangian formulation and introducing deformations in the full theory before symmetry reduction and Legendre transformations, it can be show that, in most models, the deformed covariance fails beyond spherical symmetry. Focusing, therefore, on the spherically symmetric sector, the most general deformed Hamiltonian constraint with a closed algebra is calculated. The modified Einstein equations are even explicitly solved for static solutions.\\

Inputs from quantum gravity also allow te revisit the Hawking evaporation picture. A new paradigm, advocated in \cite{Ashtekar:2020ifw}, suggests that dynamical horizons, instead of null event horizons, are what form and evaporate. During the evaporation, the trapping dynamical horizon is time-like and the marginally trapped surface shrinks in area in the future direction. Even in the semi-classical region, the interior space-time geometry develops new features when the backreaction due to the negative energy Hawking flux is taken into account. Partners of the modes that escape to $\mathcal{I}^{+}$ enter the interior of the trapping dynamical horizon and have their wavelengths enormously stretched. Based on the lessons of loop quantum cosmology, the BH singularity is expected to be resolved. In the quantum extended space-time, the singularity is replaced by a quantum ``transition surface" to the past of which there is a trapped (BH) region and to the future of which there is an anti-trapped (white-hole) region. The arguments are supported by concrete results. \\

Just at the intersection between full quantum gravity and measurements, lies the idea that gravitational wave echoes might be smoking guns for quantum horizons. It is argued in \cite{Abedi:2020ujo} that quantum black holes may be radically different from their classical counterparts in Einstein’s gravity. Among many other observables, the amplitude of the $n$-th echo can be calculated and shown to be, in tortoise coordinate,
\begin{equation}
Z_{\text{echo}}^{(n)} = {\cal T}_{\text{BH}}^{\rightarrow} {\cal R}^n ({\cal R}_{\text{BH}}^{\rightarrow})^{n-1} e^{-2 n i \tilde{\omega} r^{\ast}_0} Z_{\text{BH}} (\omega),
\end{equation}
where ${\cal R}_{\rm BH}^{\rightarrow} \equiv B_{\rm in}/B_{\rm out}$, ${\cal T}_{\rm BH}^{\rightarrow} \equiv \sqrt{\omega/ |\tilde{\omega}|} B_{\rm out}^{-1}$, $\tilde{\omega} \equiv \omega - m a / (2 M r_+)$, $r_+ \equiv M + \sqrt{M^2 - a^2}$, and the coefficients $B$ corresponds to the mode of pulsation $\tilde{\omega}$. An exhaustive account of theoretical motivations, as well as a review of the exciting and confusing state of observational searches for echoes in LIGO/Virgo data is provided. An intriguing connection between AdS/CFT and echoes is even mentioned.\\

In parallel, interesting new results were obtained on curvature invariants for charged and rotating BHs \cite{Overduin:2020aiq} and on non-singular models of magnetized BHs based on nonlinear electrodynamics \cite{Kruglov:2019wjv}.\\

Black hole physics is entering a golden age. Both on the experimental and on the theoretical sides, new results are numerous and revolutionary. To mention only a few:  Earth-like life may exist in planets orbiting supermassive BHs irradiated by their accretion disk \cite{Iorio:2019hab}, black holes might tunnel to white holes creating metastable relics \cite{Bianchi:2018mml}, etc.  
 No consensus has however yet been reached on the quantum nature of BHs, on the information paradox and not even on the number of BHs in the Galaxy. Many questions remain to be answered, from the microscopic meaning of the Bekenstein-Hawking entropy to the formation of supermassive BHs,  and the forthcoming decade should be particularly fruitful to understand better BHs but also to use them as probes for new physics. The question of the existence of primordial black holes is also a very hot one in the current context.

\bibliography{refs}

 \end{document}